\PassOptionsToPackage{hyperfootnotes=false}{hyperref}
\documentclass{article}

\usepackage{arxiv}

\usepackage{url}            
\usepackage{booktabs}       
\usepackage{amsfonts}       
\usepackage{nicefrac}       
\usepackage{microtype}      
\usepackage{microtype}
\usepackage{graphicx}
\usepackage{subfigure}
\usepackage{booktabs} 
\usepackage{amsmath, amsthm, amssymb, amsfonts}

\usepackage{hyperref}

\usepackage{siunitx}
\usepackage{amsmath}
\usepackage{mathtools}
\usepackage{cuted}
\usepackage{titlesec}
\usepackage{graphicx}
\usepackage{lastpage}
\usepackage{slashbox}
\usepackage[yyyymmdd]{datetime}
\usepackage{setspace}
\usepackage{parskip}
\usepackage{tabu}
\usepackage{multirow}
\usepackage{relsize}
\usepackage{amsmath, amsthm, amssymb, amsfonts}
\usepackage[usenames,dvipsnames,svgnames,table]{xcolor}
\usepackage{colortbl}

\definecolor{light-gray}{gray}{0.5}
\usepackage[table]{xcolor}

\usepackage{listings}
\usepackage{color}
\usepackage{algorithmic}
\usepackage{algorithm}
\usepackage{fancyhdr}
\usepackage{authblk}
\usepackage{hyperref}
\usepackage{bookmark}

\newcommand{\x}{\textbf{x}}
\newcommand{\y}{\textbf{y}}

\newcommand{\X}{\textbf{X}}
\newcommand{\Y}{\textbf{Y}}

\newcommand{\R}{\textbf{R}}

\newcommand{\CC}{\textbf{C}}

\newcommand{\Rc}{\mathcal{R}}

\newcommand*\samethanks[1][\value{footnote}]{\footnotemark[#1]}
\title{Federated Collaborative Filtering for Privacy-Preserving Personalized Recommendation System}

\author[$\dagger$]{Muhammad Ammad-ud-din}
\author[ ]{Elena Ivannikova\thanks{equal contribution}}
\author[ ]{Suleiman A. Khan\samethanks}
\author[ ]{Were Oyomno}
\author[ ]{Qiang Fu}
\author[ ]{Kuan Eeik Tan}
\author[$\dagger$]{Adrian Flanagan}
\affil[ ]{EU Cloud R\&D Center of Helsinki}
\affil[ ]{Huawei Technologies, FINLAND}

\affil[$\dagger$]{\textit {Corresponding author:{\{muhammad.amaduddin,adrian.flanagan\}}@huawei.com}}

\begin{document}
\maketitle

\begin{abstract}
	The increasing interest in user privacy is leading to new privacy preserving machine learning paradigms. In the Federated Learning paradigm, a master machine learning model is distributed to user clients, the clients use their locally stored data and model for both inference and calculating model updates. The model updates are sent back and aggregated on the server to update the master model then redistributed to the clients. In this paradigm, the user data never leaves the client, greatly enhancing the user' privacy, in contrast to the traditional paradigm of collecting, storing and processing user data on a backend server beyond the user's control. In this paper we introduce, as far as we are aware, the first federated implementation of a Collaborative Filter. The federated updates to the model are based on a stochastic gradient approach. As a classical case study in machine learning, we explore a personalized recommendation system based on users' implicit feedback and demonstrate the method's applicability to both the MovieLens and an in-house dataset. Empirical validation confirms a collaborative filter can be federated without a loss of accuracy compared to a standard implementation, hence enhancing the user's privacy in a widely used recommender application while maintaining recommender performance.

\end{abstract}

\section{INTRODUCTION}
\label{submission}

The General Data Protection Regulation (GDPR) {https://gdpr-info.eu/} in the EU (Other jurisdictions are considering similar type legislation and also an increased awareness of users of their data privacy) requires users to be both fully informed about, and consent to the collection, storage, and utilization of their personal data by other parties. GDPR changes the default option of users' personal data being harvested, stored and used to requiring explicit opt-in from the user. A default opt-in requires users to explicitly consent to the collection and use of their personal data which many fail to do. Low user opt-in rates means less data to build high performance machine learning models which in general decrease the model's performance.      

Personalized recommendation, a classical application of machine learning models~\cite{jordan2015machine} suffers badly from unreliable predictions with diverse consequences in many different domains. For example in health care, an in-accurate recommendation on treatment choices may fail the patient while in e-commerce a poor personalization service may recommend products or content of no interest to a user, resulting in a bad user experience. There exists a need to develop new machine learning methods that offer a privacy-by-design solution where we no longer need to collect and store the users' personal data, hence no longer requiring their explicit opt-in to do so, while still making the users' personal data available for building robust models.   


Thanks to advances in technology, user devices (e.g. laptops, mobile phones or tablets) have become an integral part of the machine learning process both in terms of being the source of data used in model building and the means of delivering the results of the model (e.g. recommendations) back to the user. These devices may contain user data ranging from very sensitive personal information (e.g. age, gender, location, who like what etc) to the less sensitive (e.g. downloaded applications or videos watched). 
The standard approach to model building has been to collect user data from these device and transfer it for processing to backend servers. At the servers, the data may be ``anonymized'' however, anonymization is not foolproof and can still violate user privacy when integrated with other data~\cite{sweeney2000simple}. 


To tackle this privacy problem, a Federated Learning (FL) method has been proposed recently~\cite{mcmahan2017communication}. Federated learning distributes the model learning process to the end clients (i.e. user's devices), making it possible to train a global model from user-specific local models, ensuring that the user's private data never leaves the client device enhancing the users privacy. Their proposed FL method is specific to deep learning models with use cases in image recognition and language modeling tasks. Collaborative Filtering CF (we interchangeably use the abbreviation ``CF'' for both Collaborative Filtering and Collaborative Filter) is one of the most frequently used matrix factorization models to generate personalized recommendations either independently or combined with other types of models~\cite{koren2009matrix}. 
Particularly, CF utilizes user data to build models and generate recommendations for users, in different contexts~\cite{su2009survey}. 

In this paper, we introduce the first Federated Collaborative Filter (FCF) method. We show that federation of the collaborative filter is technically challenging and formulate the updates using a stochastic gradient-based approach.
Our method aggregates user-specific gradient updates of the model weights from the clients to update the master model.
Specifically, we derive a federated version of the widely used CF method for implicit feedback datasets~\cite{Hu2008}. However, the proposed method is generalizable and can be extended to recommendation scenarios where explicit rating information is available and can also be applied to a more generic class of matrix factorization models.
We compare results between the standard and federated models on a simulated dataset as well as on two real datasets, MovieLens and an in-house service dataset. The findings confirm that the performance of the proposed FCF method is statistically similar compared to the CF model, implying no loss of accuracy while simultaneously enhancing user privacy.

\subsection{Contribution}
Our original contributions in this work are three fold: (1) we formulate the first federated collaborative filter method, (2) we demonstrate the applicability of the proposed federated method for a classical machine learning application in personalized recommendations, and (3) we empirically demonstrate that collaborative filter can be federated without loss of accuracy.

\begin{figure}[ht]
	\centering
	\includegraphics[width=\columnwidth]{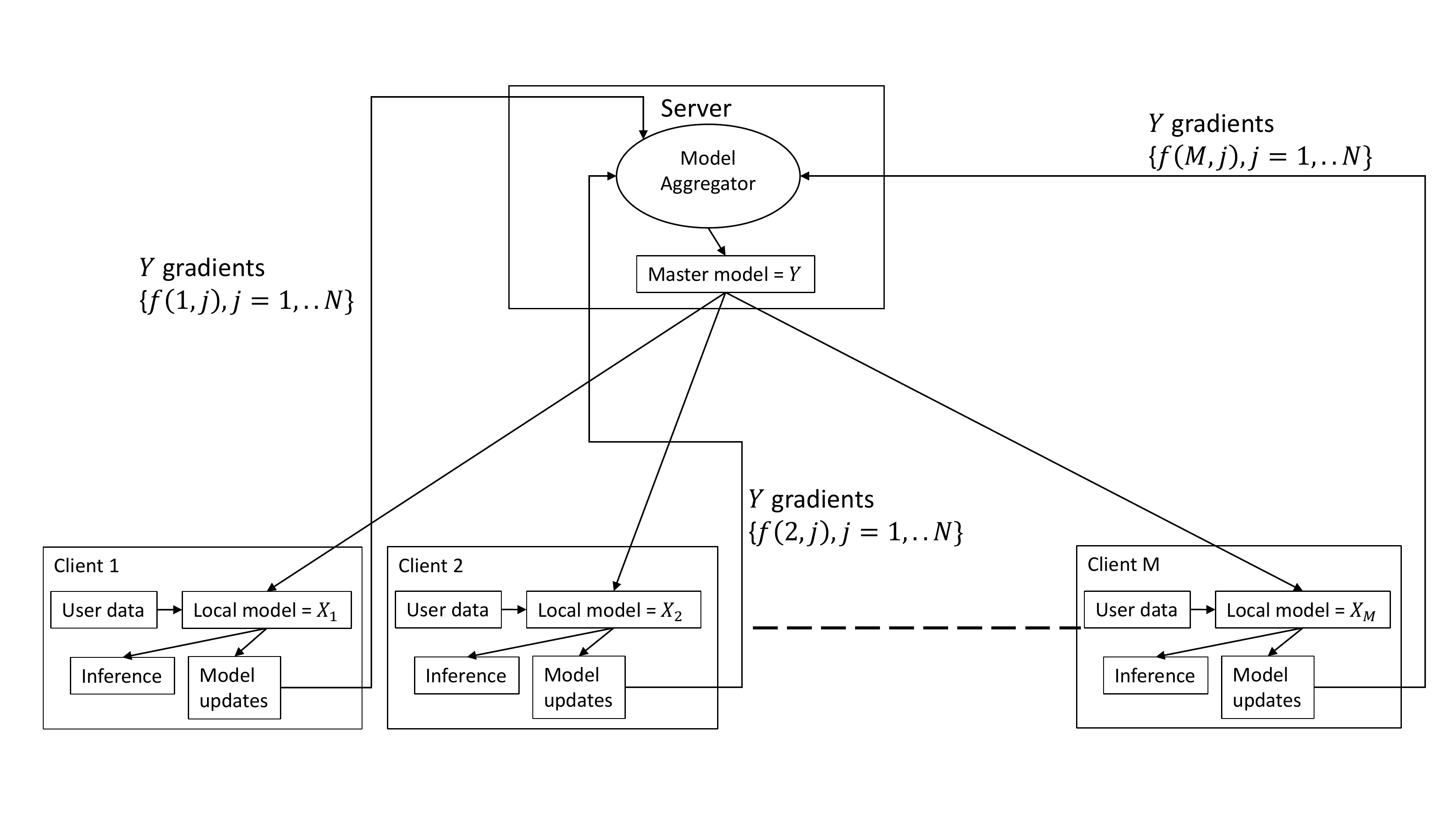}
	\label{fig:flclArch}
	\caption{Collaborative Filtering in the Federated Learning Paradigm. The Master Model $\Y$ (item-factor matrix) is updated on the server and then distributed to the clients. Each user-specific model $\X$ (user-factor matrix) remains on the local client, and is updated on the client using the local user data and $\Y$ from the server. The updates through the gradients of $\Y$ are computed on each client and transmitted to the server where they are aggregated to update the master model $\Y$.}
\end{figure}

\section{RELATED WORK}
This work lies at the intersection of three research topics: (i) matrix factorization, (ii) parallel \& distributed and, (iii) federated learning. Recently, Alternating Least Squares (ALS) and Stochastic Gradient Descent (SGD) have gained much interest and have become the most popular algorithms for matrix factorization in recommender systems~\cite{koren2009matrix}. The ALS algorithm allows learning the latent factor matrices by alternating between updates to one factor matrix while holding the other factor matrix fixed. Each iteration of an update to the latent factor matrices is referred to as an \emph{epoch}.  Although the time complexity per epoch is cubic in the number of factors, numerous studies show the ALS is well suited for parallelization~\cite{zhou2008large,takacs2009scalable, chen2011linear,schelter2013distributed,yu2014parallel}. It is not merely a coincidence that ALS is the premiere parallel matrix factorization implementation for CF in Apache Spark ( https://spark.apache.org/docs/latest/mllib-collaborative-filtering.html). 

Alternatively, SGD has become a widely adopted algorithm to implement matrix factorization for large-scale recommender applications~\cite{zinkevich2009slow,bottou2010large}. Contrary to ALS, the solution to latent factor matrices is based on a gradient descent approach, where in each iteration a small step is taken in the direction of the gradient while solving for one factor matrix and fixing the other. Consequently several iterations of gradient descent updates are required to reach the same optimum value. As compared to ALS, the SGD algorithm is efficient and simple, and the time complexity per epoch is linear in the number of factors. However, SGD requires more epochs to produce a stable model with a suitable choice of the learning rate. Furthermore, unlike ALS, parallelization of SGD is challenging, and numerous approaches have been proposed to distribute the learning of matrix factorization~\cite{zinkevich2010parallelized,recht2011hogwild,gemulla2011large,recht2013parallel}. Although these efforts clearly demonstrated the usefulness of parallelizing matrix factorization especially for large-scale recommender system applications, these work only consider the cluster and/or data center settings, and do not assume user data privacy and confidentiality that are inherent to the federated learning paradigm. We adapt the SGD algorithm to federate the seminal CF model~\cite{Hu2008}.

Federated Learning, on the other hand, a distributed learning paradigm essentially assumes user data is not available on central servers and is private and confidential.  A prominent direction of research in this domain is based on the weighted averaging of the model parameters~\cite{mcmahan2017communication,NilssonSUGJ18}. In practice, a master machine learning model is distributed to user clients. Each client updates the local copy of the model weights using the user's personal data and sends updated weights to the server which uses the weighted average of the clients local model weights to update the master model. This federated averaging approach has recently attracted much attention for deep neural networks, however, the same approach may not be applicable to a wide class of other machine learning models such as matrix factorization.  Classical studies based on federated averaging used CNNs to train on benchmark image recognition tasks~\cite{mcmahan2017communication}, and LSTM on a language modeling tasks~\cite{DBLP:journals/corr/abs-1811-03604,DBLP:journals/corr/abs-1812-02903}. As a follow-up analysis on federating deep learning models, numerous studies have been proposed addressing the optimization of the communication payloads, noisy, unbalanced~\cite{DBLP:conf/nips/SmithCST17}, non-iid and massively distributed data~\cite{DBLP:journals/corr/abs-1806-00582}.

Although our work extends the existing distributed matrix factorization and federated learning approaches, primarily we blend SGD-based distributed matrix factorization with the privacy-preserving federated learning. Compared to the commonly used federation of deep learning models, we federate the state-of-the-art CF method which is widely used in personalized recommendation systems.

\section{Collaborative Filter}
CF models the interactions between a user and a set of items. Based on the learned patterns the model recommends new items the user is expected to interact with.
In many applications, the number of users $N$ scales to several millions and the number of items $M$ to thousands. However, typically each user interacts with only a few items, leaving the user-item interaction matrix $\R \in \Rc^{N \times M}$ to be highly sparse. Low-rank factorization approaches similar to \cite{zhou2008large}, \cite{takacs2009scalable} and \cite{chen2011linear} have been shown to handle sparsity well and are able to scale the retrieval to large datasets.

Formally, the CF model is formulated as a linear combination of low-dimensional ($K$) latent factor matrices $\X \in \Rc^{K \times N}$ and $\Y \in \Rc^{K \times M}$ \cite{koren2009matrix} as
\begin{equation}
\R \sim \X^T\Y. 
\end{equation}
where  $r_{ui}$ represents the interaction between user $u$ and item $i$, for $1 \leq u\leq N, 1 \leq i \leq M$.
The interactions $r_{ui}$ are generally derived from explicit feedback such as ratings given by a user $r_{ui} \in (1, \ldots, 5)$ \cite{zhou2008large}, or implicit feedback $r_{uj}\geq1$ when the user $u$ interacted with the item $i$ and is unspecified otherwise \cite{Hu2008}. For instance, the implicit interaction could imply the user has watched a video, or bought an item from an online store, or any similar action. In this work, we consider the case of implicit feedback. 
The prediction for an unspecified $\hat{r}_{ui}$ is then given by

\begin{equation}
\hat{r}_{ui}= \x_u^T \y_i.
\label{eq:MatrixFactorisation}
\end{equation} 

The implicit feedback scenario introduces a set of binary variables $p_{ui}$ to indicate the preference of a user $u$ for an item $i$ where

\begin{equation}
p_{ui} = \begin{cases}
1 & r_{ui} > 0, \\
0 &  r_{ui} = 0
\end{cases}
\label{eq:pdef}
\end{equation}

In the implicit case, the value $r_{ui}=0$ can have multiple interpretations such as the user $u$ is not interested in the item $i$ or maybe the user is not aware of the existence of item $i$. To account for this uncertainty a confidence parameter is commonly introduced \cite{Hu2008} as

\begin{equation}
c_{ui} = 1 + \alpha r_{ui}
\label{eq:conf}
\end{equation}

where $\alpha >0$. the cost function optimizing across all users $u$ and the items $i$ over the confidence levels $c_{ui}$ is then given as

\begin{equation}
J =  \sum_u \sum_j c_{ui} (p_{ui} - \x_u^T \y_i)^2 + \lambda \Big(\sum_u \lVert \x_u \rVert^2 + \sum_i \lVert \y_i\rVert^2\Big)
\label{eq:opt}
\end{equation}

with a regularization parameter $\lambda$. 
The differential of $J$ with respect to $\x_u, \forall u$ and $\y_i, \forall i$ is given by

\begin{equation}
\frac{\partial J}{\partial \x_u} \ = \ -2 \sum_{i} \big[c_{ui} (p_{uj} - \x_u^T\y_i)\big] \y_i  + 2 \lambda \x_u.
\end{equation}

From \cite{Hu2008} the optimal solution of $\x_u = \x_{u^*}$ where $\partial J(\x_{u^*})/\partial \x_u = 0$ is defined as

\begin{equation}
\x_u^* \ = \ \big(\Y\CC^u\Y^T + \lambda I\big)^{-1}\Y\CC^u p(u),
\label{eq:x_opt}
\end{equation}

where $\CC^u \in \Rc^{N \times N}$ is a diagonal matrix with $\CC^u_{ii} = c_{ui}$ and $p(u) \in \Rc^{N \times 1}$ contains the $p_{ui}$ values for the user $u$. Similarly for $y_i$ 

\begin{equation}
\frac{\partial J}{\partial \y_i} \ = \ -2 \sum_{u} \big[c_{ui} (p_{ui} - \x_u^T\y_i)\big] \x_u  + 2 \lambda \y_i,
\label{eq:ydiff}
\end{equation}

\begin{equation}
\y_i^* \ = \ \big(\X\CC^i\X^T + \lambda I\big)^{-1} \X\CC^ip(i)
\label{eq:y_opt}
\end{equation}

%


where $\CC^i \in \Rc^{M \times M}$ is a diagonal matrix with $\CC^i_{uu} = c_{ui}$ and $p(i) \in \Rc^{M \times 1}$ contains the $p_{ui}$ values for item $i$. 
Using Eqs.~\ref{eq:x_opt} and \ref{eq:y_opt} \cite{Hu2008} describe a computationally efficient Alternating Least Squares (ALS) algorithm to find the optimum values of $\X, \Y$. 
The ALS algorithm solves for the $\X, \Y$ factor matrices by alternating between updates. Multiple \emph{epochs} are carried out to update $\X, \Y$, until a suitable convergence criteria is satisfied.

\section{Federated Collaborative Filtering}




We now introduce the Federated Collaborative Filter (FCF) Algorithm \ref{alg:fcf} that extends the CF model to the federated mode. 
The key idea is to carry out model updates such that the user's private interaction data is not transferred to the server.
The FCF distributes parts of the model computation to the clients and is illustrated in Figure~\ref{fig:flclArch}. Specifically, the method defines three core components as below.

\begin{enumerate}
	\item All the item factor vectors $y_i, i= 1, \ldots, M$ are updated on the server and then distributed to each client $u$.
	\item The user factor vectors $x_u, u \in \{1, \ldots, N\}$ are updated locally on the client $u$, using the user $u$'s own data and the $y_i, i = 1, \ldots, M$ from the server.
	\item The updates through the gradients $\delta y_{u i}$ are calculated for the item $i$ on each client $u$ and transmitted to the server where the gradients are aggregated to update $y_i$. This is in contrast to the existing federated learning architectures \cite{McMahan17}, where the clients directly compute the updates of the parameters $y_{u i}$ which are then aggregated on the server to update the master model.
\end{enumerate}

We next discuss the federated model updates in detail.
\subsection{Federated User Factor Update}
In each update iteration, the server sends the latest item factor vectors $\{\y_i, i=1, \ldots, M\}$ to each client. The user's own data $r_{ui}$ is used to compute $p(u)$ and $\CC^u$ using Eq.~\ref{eq:pdef} and \ref{eq:conf}, respectively. The method then uses Eq.~\ref{eq:x_opt} to update the $\x_u^*$ locally for each client. The updates can be carried out independently for each user $u$ without reference to any other user's data.

\subsection{Federated Item Factor Update}
To update the item factor vectors using Eq.~\ref{eq:y_opt}, the user factor vectors $\x_i$ $\forall i \in \{1, \ldots, N\}$ and the interaction of the users with items $i$ through the $\CC^u$ and $p(u)$ are required. Therefore, the updates of the $\y_i$ cannot be done on the clients and must be carried out on the master server. However, in order to preserve the user's privacy the user-item interactions should remain on the client's own device only, therefore, Eq.~\ref{eq:y_opt} cannot be used for computing $\y_i^*$. This is the fundamental problem when federating CF and we present the first solution to address this challenge.

We next adapt and demonstrate a stochastic gradient descent approach to allow for the update of the $\y_i$ vectors on the server, while preserving the user's privacy.

Formally, $y_i$ is updated on the master server as

\begin{equation}
\y_i \ = \ \y_i \ - \ \gamma \frac{\partial J}{\partial \y_i},
\label{eq:gradientDescent}
\end{equation}

for some gain parameter $\gamma$ to be determined and $\partial J/\partial \y_i$ as given in Eq.~\ref{eq:ydiff}. However, Eq.~\ref{eq:ydiff} contains a component which is a summation over all users $u$. We therefore, define $f(u, i)$ as

\begin{equation}
f(u, i) \ = \ \big[c_{ui} (p_{ui} - \x_u^T\y_i)\big] \x_u,
\label{eq:fdiff}
\end{equation}

where $f(u, i)$ is calculated on each client $u$ independently of all the other clients. All the clients then report back the gradient values $f(u, i), \ i=\{1,\ldots, N\}$ to the master server.
As a result, Eq.~\ref{eq:ydiff} can then be re-written as an aggregation of the client gradients and computed on the master as

\begin{equation}
\frac{\partial J}{\partial \y_i} \ = \ -2 \sum_{u} f(u, i)  + 2 \lambda \y_i.
\label{eq:sumf}
\end{equation} 


Finally, the item specific gradient updates in Eq.~\ref{eq:sumf} are then used to update $\y_i$ on the master server using Eq.~\ref{eq:gradientDescent}.


The proposed FCF method alternates between solving for $\X$ using Eq.~\ref{eq:x_opt} and then solving for $\Y$ using Eq.~\ref{eq:gradientDescent}. However, as $\Y$ are updated using a gradient descent approach, multiple iterations of gradient descent updates are required to reach the optimum value of $\Y$. Therefore, a single epoch of FCF consists of an update to $\X$ as in CF and then several gradient descent steps to update $\Y$. 

We analyze the Adaptive Moment Estimation (Adam) method \cite{kingma2015adam} for FCF. In Adam the gradient descent is carried out in two separate parts which record an exponentially decaying average of past gradients $m_t$ and squared gradients $v_t$,
\begin{equation}
m \ = \ \beta_1 m \ + \ (1 - \beta_1) \frac{\partial J}{\partial \y_i} 
\label{eq:adam_m}
\end{equation}
and 
\begin{equation}
v \ = \ \beta_2 v \ + \ (1 - \beta_2) \Bigg(\frac{\partial J}{\partial \y_i}\Bigg)^2 
\label{eq:adam_v}
\end{equation} 
with $0 < \beta_1, \beta_2 < 1$. Typically, $m$ and $v$ are initialized to $0$ values and hence biased towards $0$. To counteract these biases, corrected versions of $m, v$ are given by
\begin{equation}
\hat{m} = \frac{m}{1 - \beta_1},
\end{equation}
and
\begin{equation}
\hat{v} = \frac{v}{1 - \beta_2}.
\end{equation}
The updates are then given by
\begin{equation}
\y_i \ = \ \y_i - \frac{\gamma}{\sqrt{\hat{v}} + \epsilon}\hat{m}
\end{equation}
where $0 < \gamma < 1.0$ is a constant learning rate and $0 < \epsilon \ll 1$ e.g. $10^{-8}$ to avoid a divide by $0$ scenario. 

\begin{algorithm}[b]
	\caption{Federated Collaborative Filter (FCF)}
	\label{alg:fcf}
	\textbf{input}: M clients
	\begin{algorithmic}
		\STATE FL Server
		\STATE {\bfseries Initialize:} $\Y$
		\FOR{each client $m \in M$ in parallel}
		\STATE $\nabla \Y^{(m)}$ = ClientUpdate($\Y$)
		\STATE $\Y$ = $\Y - \gamma \sum_{m} \nabla \Y^{(m)}$ using Eqs.~\ref{eq:gradientDescent} \& \ref{eq:sumf}
		\ENDFOR
		\STATE FL Client
		\STATE {\bfseries function:} ClientUpdate($\Y$) 
		\STATE update user factor $\x_u$ using Eq.~\ref{eq:x_opt}
		\STATE compute item factor $\Y^{(m)}$ gradients: $\nabla \Y^{(m)}$ using Eq.~\ref{eq:fdiff}
		\STATE return $\nabla \Y^{(m)}$ to server
	\end{algorithmic}
\end{algorithm}

\subsection{Privacy-by-Design Solution}
Our privacy preserving federated solution does not requires the user's identity to be known at the server. This is primarily because each user sends the updates for $f(u, i)$ which are aggregated in Eq.~\ref{eq:sumf} without reference to the user's identity. 

\section{DATA}
\subsection{Simulated data}
A simulated dataset of user-item interactions was generated by randomly simulating users, movies, and view events. Specifically, we created a user-item interaction matrix consisting of zeros and ones. 
The data was required to be 80\% sparse with the additional constraints that each user has at least eight view events and each item was watched at least once. The data dimensionality is specified in Table \ref{table:datasets}.

\subsection{MovieLens}
The MovieLens rating datasets have been collected and made available by the GroupLens research group (https://grouplens.org/). The dataset~\cite{harper2016movielens} used in this research work contains 1,000,209 anonymous ratings of 3952 movies made by 6040 users who joined MovieLens in the year of 2000. The data contains user-item interactions characterized by the user id, movie id, and explicit rating information. We transform the explicit ratings to the implicit feedback scenario of our proposed approach. Specifically, we assume that users watched the videos they rated and are not aware of the remaining ones. As a result, the explicit rating wherever available irrespective to its value, is transformed to one while missing ratings are assigned zeros. The transformation, therefore, represents the real implicit feedback data.


\begin{table}[ht]
	\centering
	\resizebox{0.5\columnwidth}{!}{%
		\centering
		\begin{tabular}{l|ccc}
			\toprule
			Dataset & \# users & \# movies & \# view events\\
			\midrule
			Simulated & 5000 & 40  & 8 \\
			MovieLens & 6040 & 3952 & 20\\
			In-house Production Data & 6077 & 1543 & 20\\
			\bottomrule
	\end{tabular}}
	\caption{Overview of the datasets used in the study, where \# view events refers to the minimal number of videos each user watched.}
	\label{table:datasets}
\end{table}



\subsection{In-house Production Dataset}
The in-house production dataset consists of a data snapshot extracted from the database that contains user view events. 
We carried out experiments using a subset of the anonymized dataset that was created by selecting view events for users who watched more than 20 videos and for videos that were viewed by more than 100 users. 
The overview of the dataset is listed in Table \ref{table:datasets}.

\section{EXPERIMENTS AND RESULTS}
\subsection{Experimental setup}

The CF and FCF models have three hyper-parameters in common: $\alpha$, $\lambda$ and the number of factors $K$. Whereas, the hyper-parameters related to the learning rate are specific to FCF only.
To chose optimal hyper-parameters on real datasets, we used five-fold cross validation using Bayesian optimization approach~\cite{snoek2012practical}. We specified bounds for $K \in [2,4]$, $\alpha \ \& \ \lambda \in [1,5]$ and learned the optimal hyper-parameters using the python implementation\footnote[6]{https://github.com/fmfn/BayesianOptimization}. We used a standard CF model to select the common hyper-parameters, whereas the parameters related to the Adam's learning rate were optimally chosen using the FCF model. Similar to CF, we used the Bayesian optimization to infer the FCF specific hyper-parameters using the bounds for $\beta_1\ \&\ \eta \in [0.0999,0.50], \beta_2 \in [0.0999,0.80], \epsilon \in [1e-8,1e-5]$. The number of gradient descent iterations per epoch was set to $10$.

For the simulated data experiments, we also wanted to evaluate how varying several hyper-parameters effect the model's behavior. Therefore, these were varied during the experiments. However, when not specified, the default values used for hyper-parameters in simulated data experiments were $\alpha=1.0$, $\lambda=1.0$, $\gamma=0.05$, $K = 4$ and the number of gradient descent iterations per epoch as $20$.
All experiments were run for $20$ epochs.
\newpage
\subsection{Convergence Analysis}
\begin{figure*}[ht]
	\begin{minipage}{.33\textwidth}
		\centering
		\includegraphics[width=\textwidth,trim=0cm 5cm 0cm 5cm]{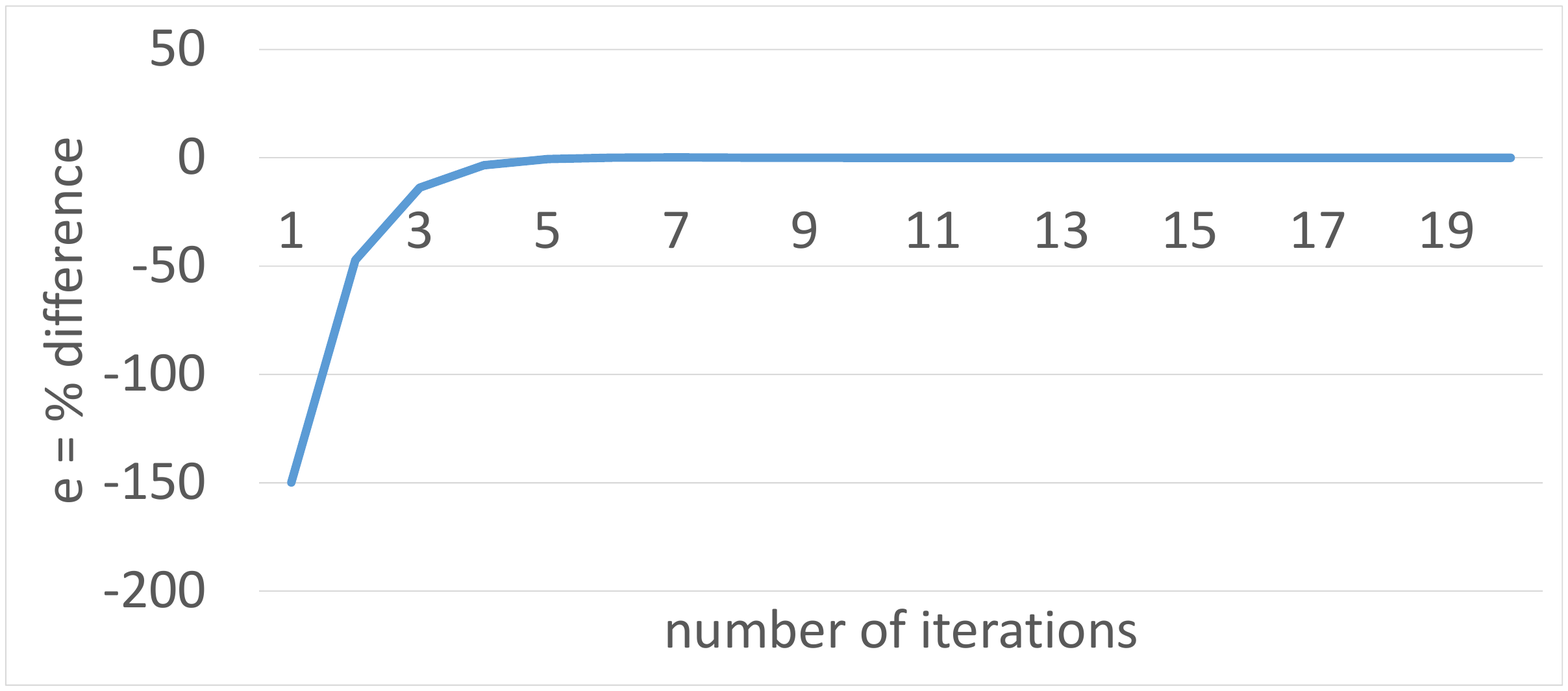}
	\end{minipage}
	\begin{minipage}{.33\textwidth}
		\centering
		\includegraphics[width=\textwidth,trim=0cm 5cm 0cm 5cm]{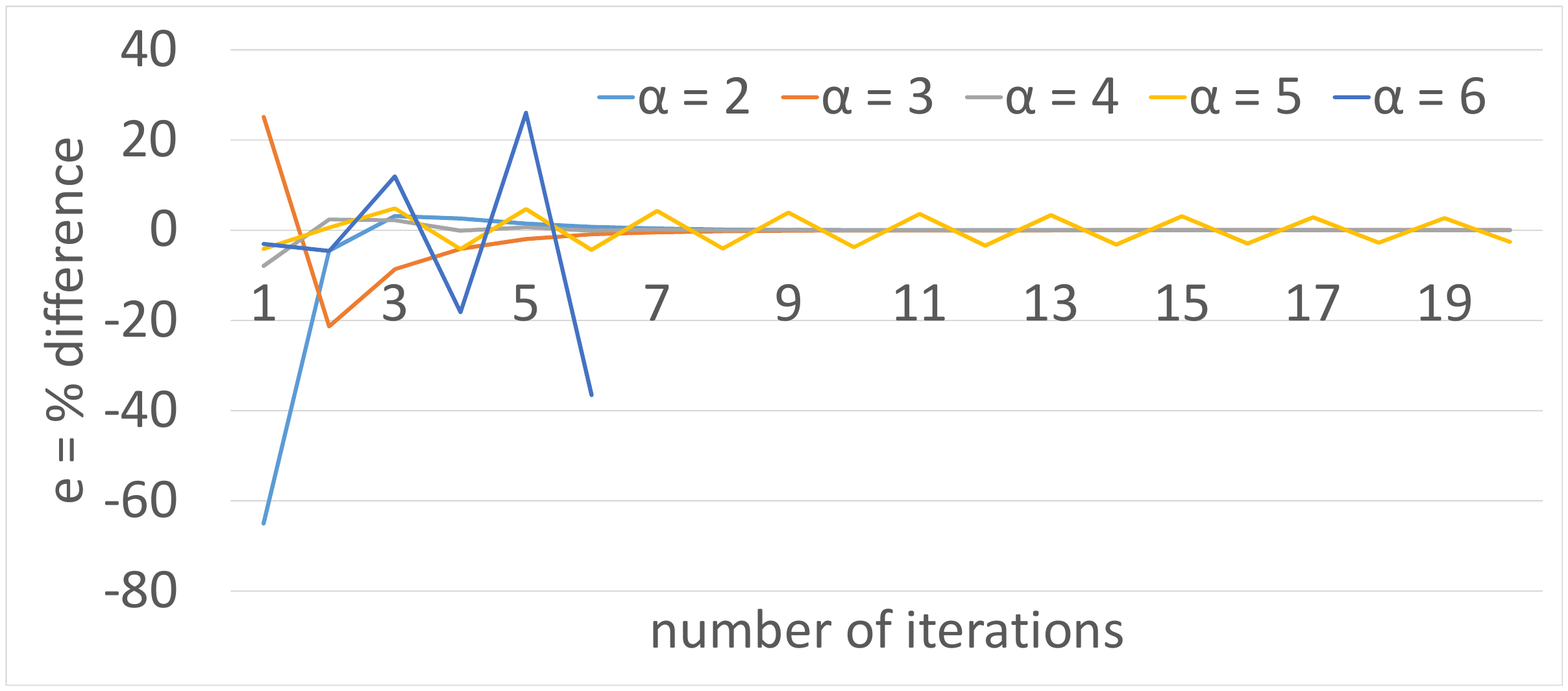}
	\end{minipage}
	\begin{minipage}{.33\textwidth}
		\centering
		\includegraphics[width=\textwidth,trim=0cm 5cm 0cm 5cm]{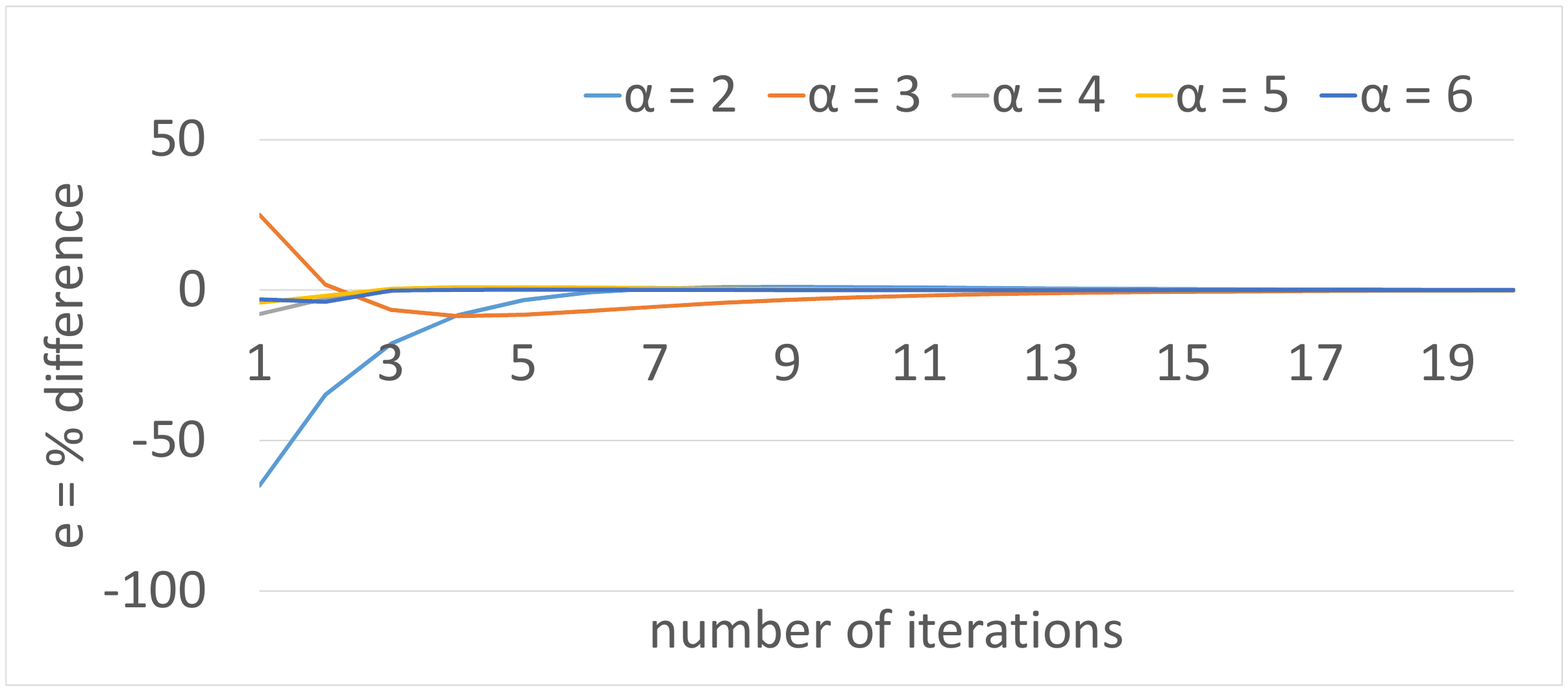}
	\end{minipage}
\end{figure*}
\begin{figure*}[ht]
	\begin{minipage}{.33\textwidth}
		\centering
		\includegraphics[width=\textwidth,trim=0cm 5cm 0cm 11cm]{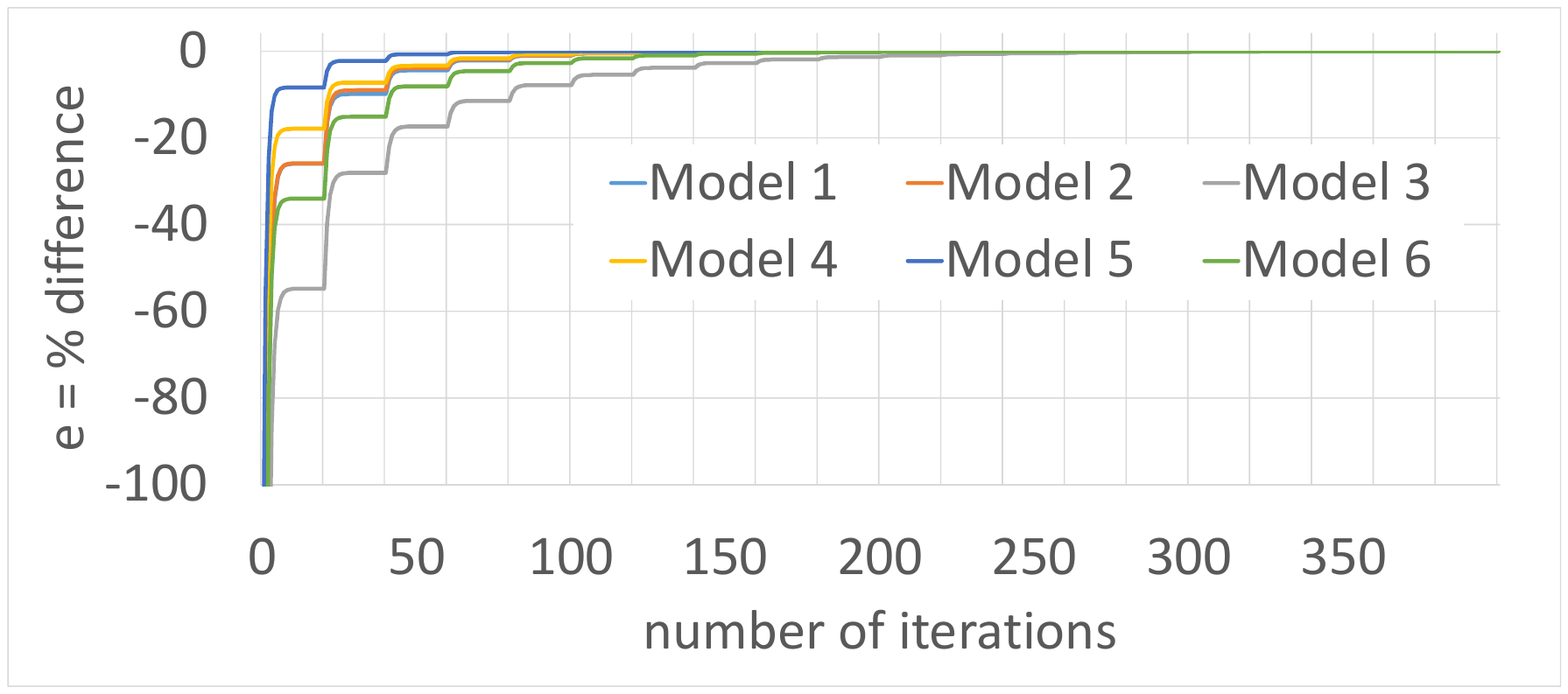}
	\end{minipage}
	\begin{minipage}{.33\textwidth}
		\centering
		\includegraphics[width=\textwidth,trim=0cm 5cm 0cm 11cm]{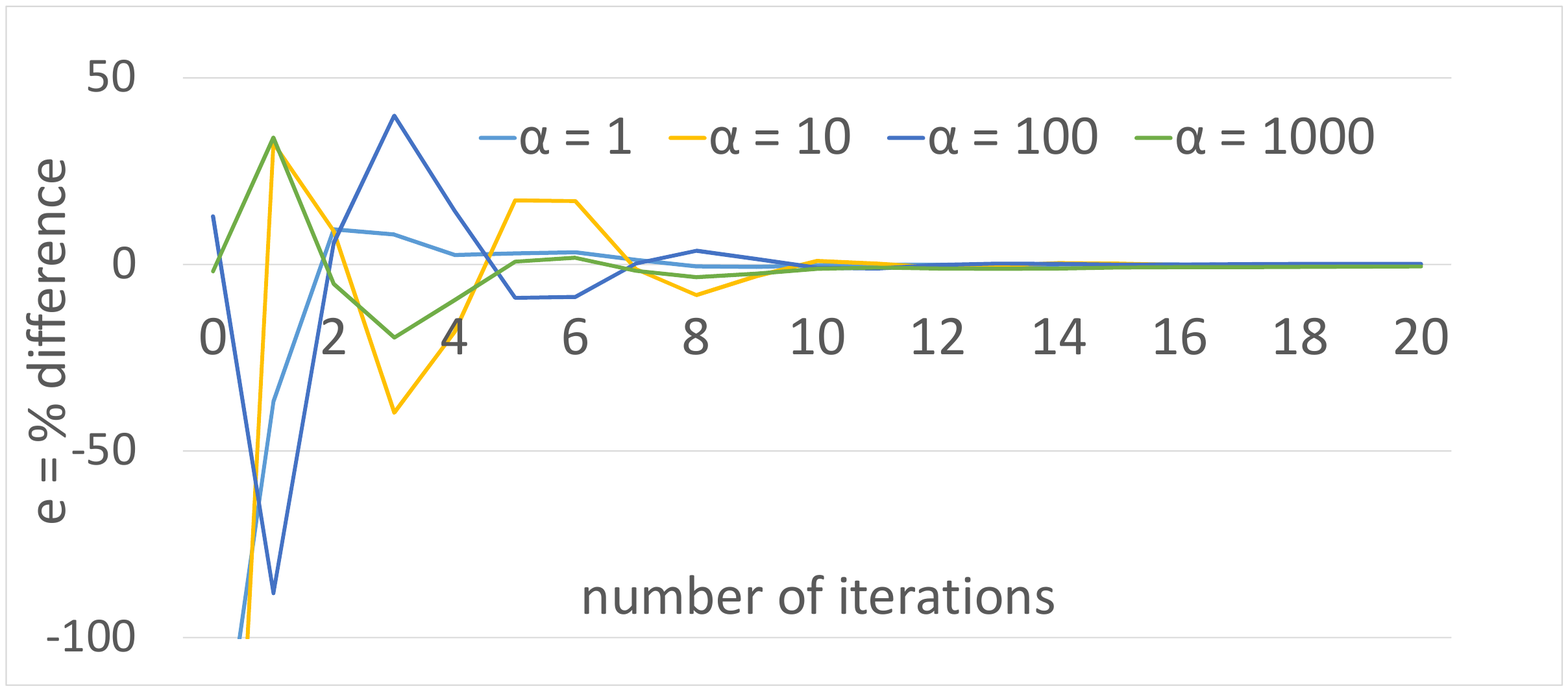}
	\end{minipage}
	\begin{minipage}{.33\textwidth}
		\centering
		\includegraphics[width=\textwidth,trim=0cm 5cm 0cm 11cm]{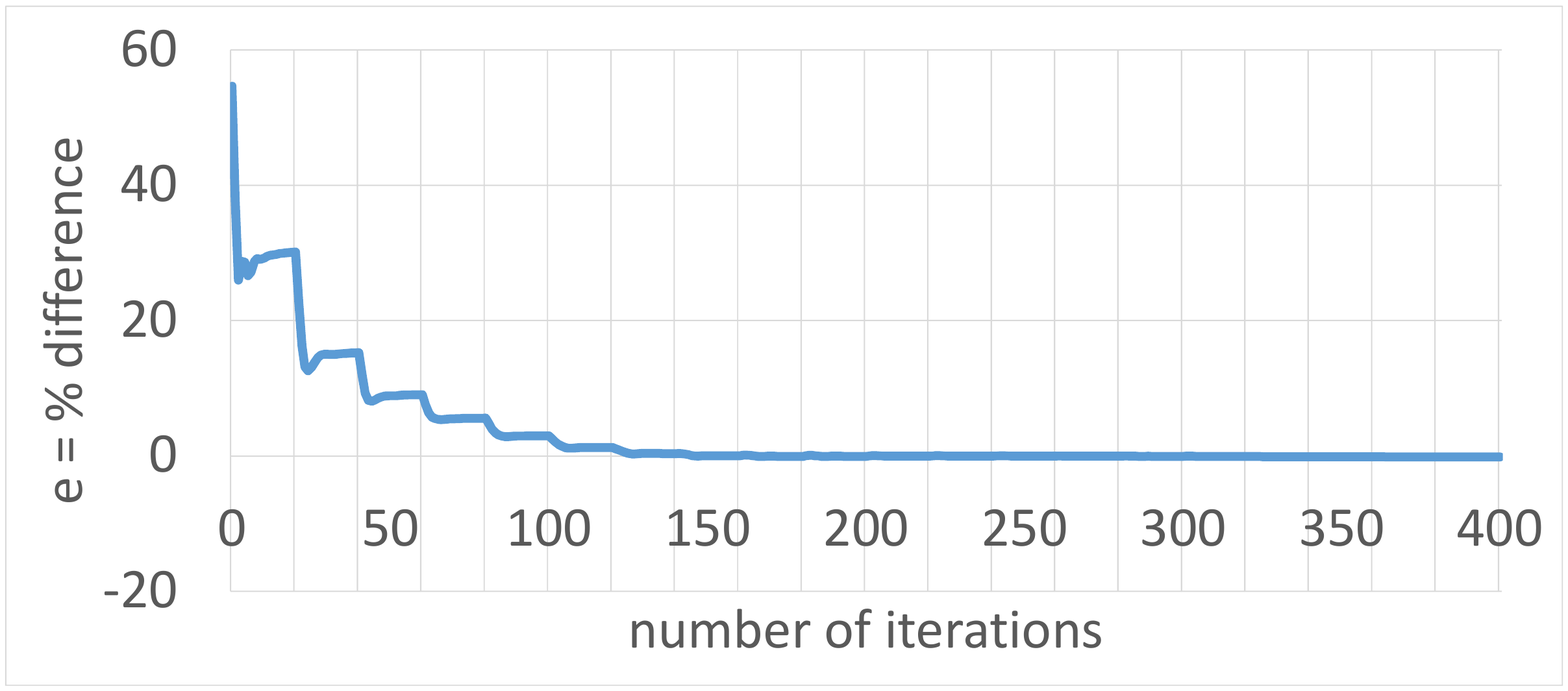}
	\end{minipage}
	\vskip -0.4in
	\caption{
		Convergence analysis of FCF model. The y-axis represents the $\%$ difference between the elements of latent factors $\Y_{\text{FCF}}$ to $\Y_{\text{CF}}$ whereas the x-axis shows iterations. With 20 gradient descent iterations per epoch the vertical lines indicate the start and end of an epoch.
		\textbf{Top-Left}, $\alpha = 1, \gamma = 0.05$, epoch $= 1$;
		\textbf{Top-Middle}, $\alpha = 2-6, \gamma = 0.05$, epoch $= 1$;
		\textbf{Top-Right}, $\alpha = 2-6, \gamma = 0.025$, epoch $= 2$;
		\textbf{Bottom-Left}, $\alpha = 10, \gamma = 0.05$, epoch $= 20$;
		\textbf{Bottom-Middle}, $\alpha = 1-1000, \gamma = 0.2$, epoch $= 1$ using Adam adaptive learning rate;
		\textbf{Bottom-Right}, $\alpha = 10, \gamma = 0.2$, epoch $= 20$ using Adam adaptive learning rate.
	}
	\label{fig:conv_analysis}
\end{figure*}


The proposed method to ``federating" the CF can be seen as a variation of the standard ALS algorithm. The ALS algorithm learns the $\X, \Y$ factors by alternating between updates to $\X, \Y$ using Eq.~\ref{eq:x_opt} and \ref{eq:y_opt} respectively. Each iteration of an update to both $\X, \Y$ is referred to as an \emph{epoch}. In CF, each ALS alteration of the updates between $\X, \Y$ is convex, hence the rate of convergence of the model is expected to be quick and consistent.

Our proposed FCF method uses a similar approach but instead alternates between updating $\X$ using Eq.~\ref{eq:x_opt} and then solving for $\Y$ using gradient descent Eq.~\ref{eq:ydiff}. The update of $\Y$, therefore, requires several iterations of gradient descent to reach the optimum value.  Naturally, the FCF is also run in epochs where each epoch consists of an update to $\X$ as in CF and then several gradient descent steps to update $\Y$. 

We demonstrate that FCF and CF models converge to the same optimum. 
The solutions are compared by calculating the average percentage difference between the elements of latent factors as
\begin{equation}
e \ =  \frac{ \sum_i \sum_k (\Y_{\text{CF}}(i, k) - \Y_{\text{FCF}}(i, k) ) /\Y_{\text{CF}}(i, k)}{M \times K} \times 100 \nonumber
\end{equation}    

where $Y_{\text{FCF}}$ is the item-factor matrix of FCF and $Y_{\text{CF}}$ is the item-factor matrix of CF.

First we compare the results after one epoch of the CF and FCF training, with both initialized with the same $\X, \Y$. To ensure consistency, $\X$ is updated in both cases resulting in the same values of $\X$ for both algorithms initially. The $\Y_{FCF}$ is updated using Eq.~\ref{eq:gradientDescent} and compared to the learned $\Y_{CF}$. We found that the FCF model converges after $5$ gradient descent iterations to $\approx 0\%$ difference between the $\Y_{CF}$ and $\Y_{FCF}$ (Figure \ref{fig:conv_analysis}, top-left). The result confirms that the two models convergence to the same optimal. This is typically the case in any form of iterative optimization process. However, it does indicate that the FCF algorithm converges to the optimum ALS solution for this simple scenario.

Typically, the optimization of gradient descent approach depends on the appropriate choice of learning rate parameter $\gamma$ ~\cite{zinkevich2010parallelized}, therefore, we next investigated the effects of $\gamma$ on varying the value of implicit confidence parameter $\alpha$ (in Eq.~\ref{eq:conf}). We observed that with $\gamma=0.05$, stable convergence of the $\Y_{FCF}$ factor matrices was achieved for smaller values of $\alpha \in \{1, 2, 3, 4\}$, however, the convergence became un-stable for the larger $\alpha$ values (Figure \ref{fig:conv_analysis}, top-middle). However, decreasing the value of $\gamma$ by half, results in a stable convergence for all the values of $\alpha$ (Figure \ref{fig:conv_analysis}, top-right).


We next compared the model parameters by increasing the number of training epochs from $1$ to $20$ to evaluate if the gradient descent estimation of $\Y_{\text{FCF}}$ converges globally, for higher values of $\alpha = 10$.
To obtain robust evaluation, for FCF, we trained 6 models with the same hyper-parameters but different initialization of the factors $\X_{\text{FCF}}, \Y_{\text{FCF}}$, and evaluated the percentage difference at each gradient descent step of each epoch.

The results demonstrate that all 6 FCF model runs converged to the optimal solution of CF as shown in Figure \ref{fig:conv_analysis}, bottom-left, although at different rates.
Specifically, 5 out of the 6 model runs converged within $6$ epochs to the final $ \Y_{\text{CF}}$ solution. These simulations indicate that the convergence is quite robust although the rates of convergence can vary. For consistency purpose, we repeated the experiments with a higher $\gamma=0.1$ (Supplementary Figure S1), which also led to convergence of all the models, however, required more iterations.


Several approaches can be used to stabilize the gradient descent methods including adaptively changing the learning rate $\gamma$ or introducing gradient checking~\cite{kingma2015adam,DBLP:journals/corr/Ruder16}.
Therefore, to tackle the diverging gradients, we introduce an adaptive learning rate based on Adam optimizer~\cite{kingma2015adam} into the FCF model. Using cross-validation, we identified values of $\beta_1 = 0.4$, $\beta_2 = 0.99$ and $\gamma = 0.2$ to produce stable convergence, where $\beta_1$ and $\beta_2$ are as defined in Eq.~\ref{eq:adam_m} and \ref{eq:adam_v}. The convergence for a single epoch and different values of $\alpha$ ranging form $1-1000$ are shown in Figure \ref{fig:conv_analysis}, bottom-middle. The results confirm that the convergence is stable across several orders of magnitude of the confidence parameter $\alpha$, as the algorithm convergences around $\sim10$ iterations in all cases. Finally, we also analyzed the adaptation of Adam's learning rate on the model's convergence over $20$ epochs with  $\alpha = 10$ and a substantially high $\gamma = 0.2$  confirming that the model converged to the CF solution within $6-7$ epochs (Figure \ref{fig:conv_analysis}, bottom-right).


In summary, the comprehensive simulations confirm that FCF models converged to the CF model's optimal solution, although at different rates. importantly, the Adam's learning rate is needed to stabilize the convergence in the presence of increasing implicit confidence parameter $\alpha$.

\subsection{Recommendation Performance}
We next evaluate the recommendation performance of our proposed Federated Collaborative Filter in comparison to the standard Non-Federated Collaborative Filter \cite{Hu2008}. 
To comprehensively compare the performance metrics we use the standard evaluation metrics \cite{BOBADILLA2013109}, \textbf{Precision}, \textbf{Recall}, \textbf{F1}, and \textbf{Mean Average Precision (MAP)} and \textbf{Root Means Square Error (RMSE)} for the top $10$ predicted recommendations. For completeness the metrics are defined in Supplementary Material. We also compute 
``$\textbf{diff \%}$" between performance metrics of FCF and CF as
\begin{equation}
\textbf{diff \%} = \lvert{\frac{\text{MetricMean(FCF)} - \text{MetricMean(CF)}}{\text{MetricMean(CF)}}}\rvert \times 100 \nonumber
\label{eqn:diff}
\end{equation}

\begin{table}[ht]
	\centering
	\resizebox{0.5\columnwidth}{!}{%
		\begin{tabular}{@{}cccc@{}}
			\toprule
			& CF & FCF & diff \% \\ \midrule
			\multicolumn{4}{c}{Movie-Lens} \\
			Precision & 0.3008 $\pm$ 0.0079 & 0.2993 $\pm$ 0.0083 & 0.4987 \\
			Recall & 0.1342 $\pm$ 0.0044 & 0.134 $\pm$ 0.0046 & 0.149 \\
			F1 & 0.1552 $\pm$ 0.0047 & 0.1548 $\pm$ 0.0049 &  0.2577 \\
			MAP & 0.2175 $\pm$ 0.008 & 0.2155 $\pm$ 0.0082  & 0.9195 \\
			RMSE & 0.6988 $\pm$ 0.056 & 0.6994 $\pm$ 0.0558 & 0.0859 \\ \midrule
			\multicolumn{4}{c}{In-House} \\
			Precision & 0.0916 $\pm$ 0.0173 & 0.0914 $\pm$ 0.0172 & 0.2183  \\
			Recall & 0.1465 $\pm$ 0.0289 & 0.146 $\pm$ 0.0289 & 0.3413  \\
			F1 & 0.1104 $\pm$ 0.0214 & 0.11 $\pm$ 0.0213 & 0.3623  \\
			MAP & 0.0669 $\pm$ 0.017 & 0.0664 $\pm$ 0.0171 & 0.7474  \\
			RMSE & 0.8076 $\pm$ 0.0316 & 0.8083 $\pm$ 0.0323 & 0.0867 \\ \midrule
			\multicolumn{4}{c}{Simulated} \\
			Precision & 0.2014 $\pm$ 0.0057 & 0.2013 $\pm$ 0.0059  & 0.0497 \\
			Recall & 0.8867 $\pm$ 0.0196 & 0.8863 $\pm$ 0.0199 & 0.0451 \\
			F1 & 0.3208 $\pm$ 0.0088 & 0.3207 $\pm$ 0.009 & 0.0312 \\
			MAP & 0.5805 $\pm$ 0.0341 & 0.5798 $\pm$ 0.0341 & 0.1206 \\
			RMSE & 0.5387 $\pm$ 0.0193 & 0.5391 $\pm$ 0.0192 & 0.0743 \\ \bottomrule			
		\end{tabular}
	}
	\caption{Comparison of the test set performance metrics between Collaborative Filter (CF) and Federated Collaborative Filter (FCF) using different metrics averaged over all users. The values denote the \textbf{mean} $\pm$ \textbf{standard deviation} across 10 different model builds. The \textbf{diff \%} refers to the percentage difference between the \textbf{mean} values of CF and FCF.}
	\label{table:als_fcf_metrics}
\end{table}

Table \ref{table:als_fcf_metrics} shows test set performance metrics averaged across users over 10 rounds of the model rebuilds for the two real datasets and simulated dataset. In each model build, the data for each user was randomly divided into 60\% training, 20\% validation and 20\% test sets. While validation and training sets were used to find optimal hyper-parameters and to learn model parameters, for example, the latent factors, the test set was purely employed to predict recommendations and to evaluate the performance scores on unseen user data. 
The results confirm that the FCF and CF model results are very similar in terms of test set recommendation performance metrics. On average, the percentage difference \textbf{diff \%} CF and FCF across any of the five metrics is less than $0.5\%$. The standard deviations \textbf{std} are also small suggesting the multiple runs converge to stable and comparable solutions.
\begin{figure*}[ht]
	\begin{minipage}{.19\textwidth}
		\centering
		\includegraphics[width=\textwidth]{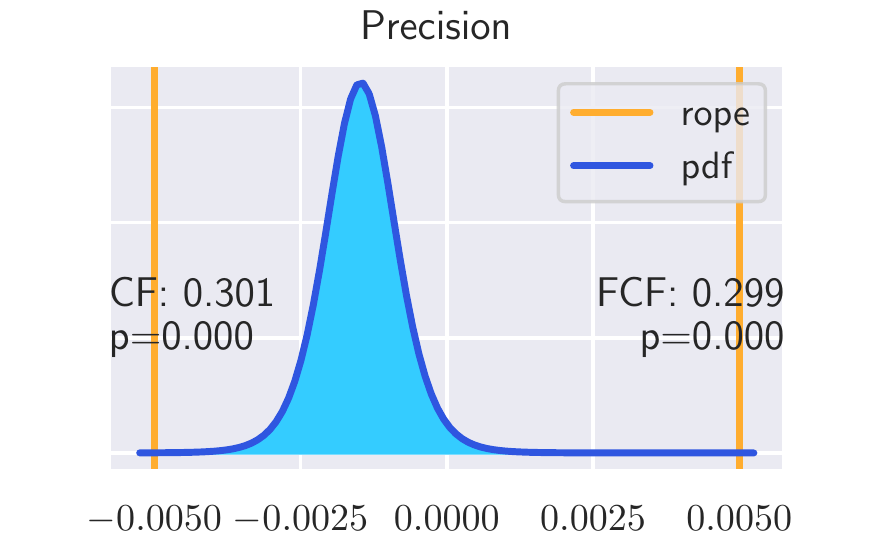}
	\end{minipage}
	\begin{minipage}{.19\textwidth}
		\centering
		\includegraphics[width=\textwidth]{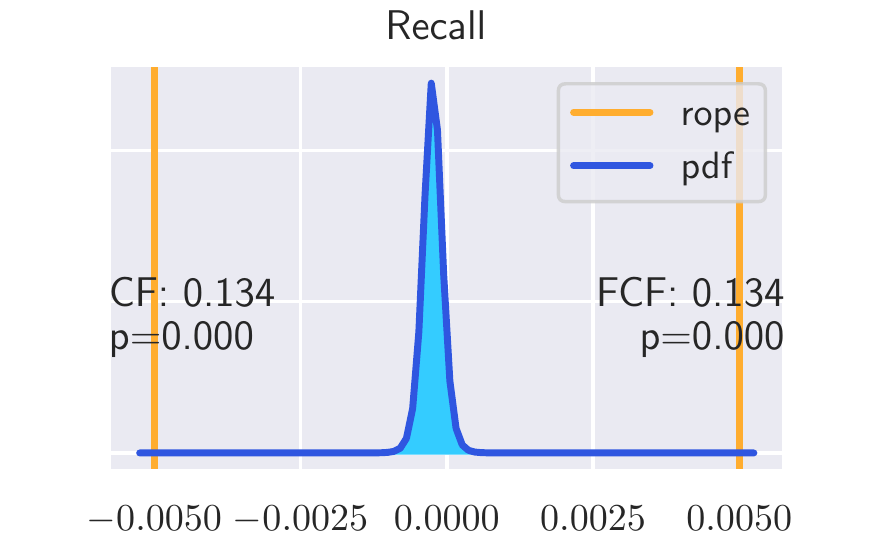}
	\end{minipage}
	\begin{minipage}{.19\textwidth}
		\centering
		\includegraphics[width=\textwidth]{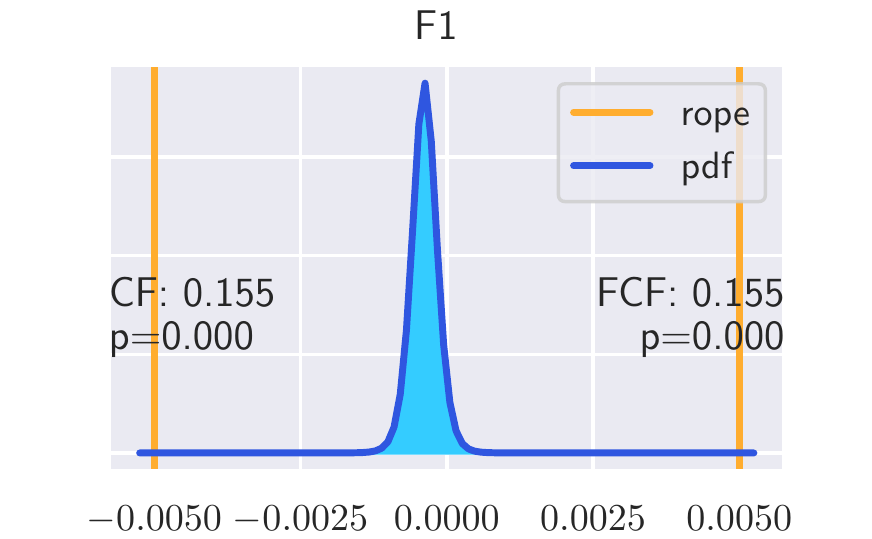}
	\end{minipage}
	\begin{minipage}{.19\textwidth}
		\centering
		\includegraphics[width=\textwidth]{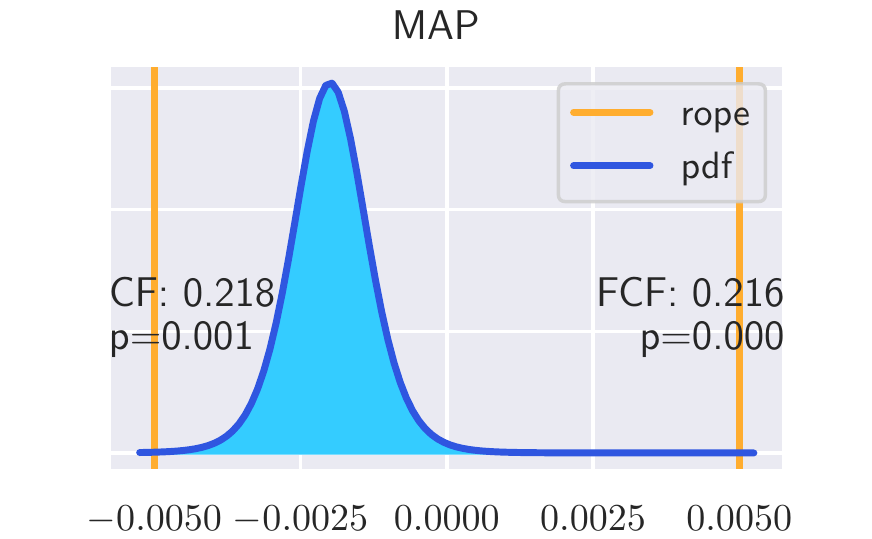}
	\end{minipage}
	\begin{minipage}{.19\textwidth}
		\centering
		\includegraphics[width=\textwidth]{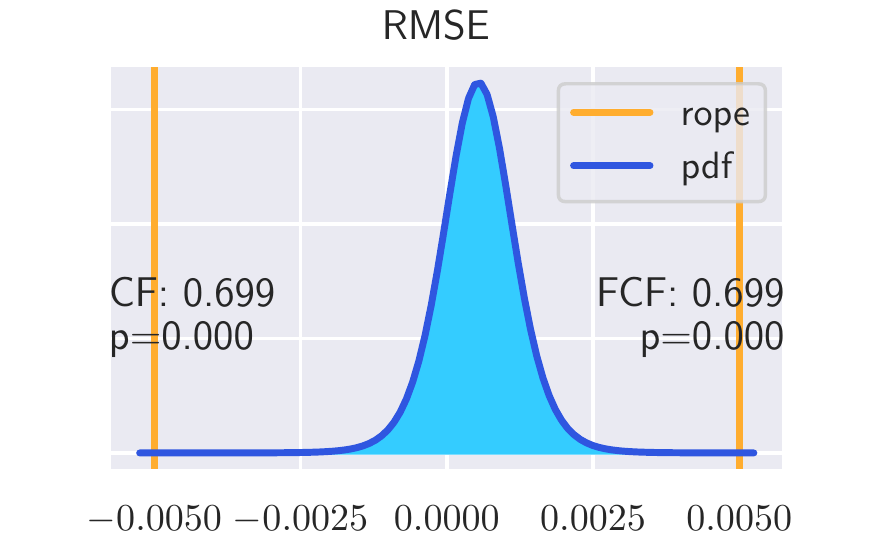}
	\end{minipage}
\end{figure*}
\begin{figure*}[ht]
	\begin{minipage}{.19\textwidth}
		\centering
		\includegraphics[width=\textwidth]{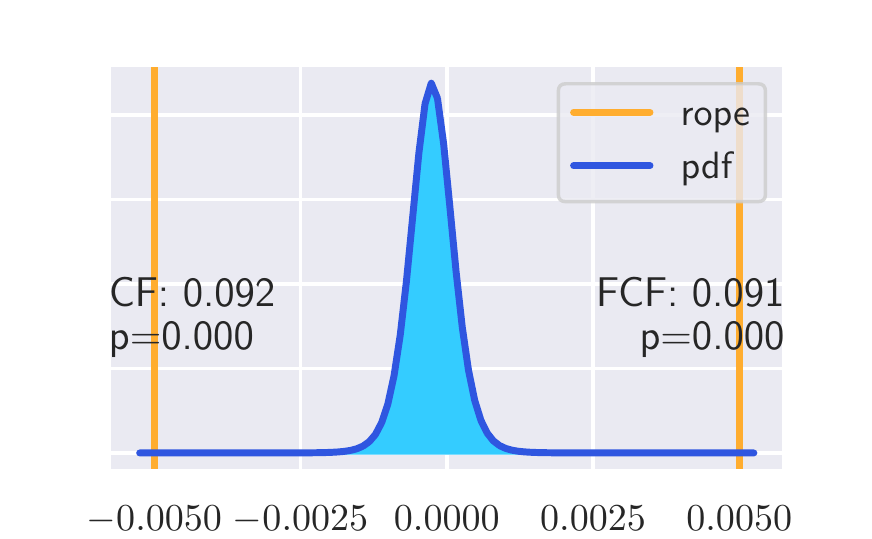}
	\end{minipage}
	\begin{minipage}{.19\textwidth}
		\centering
		\includegraphics[width=\textwidth]{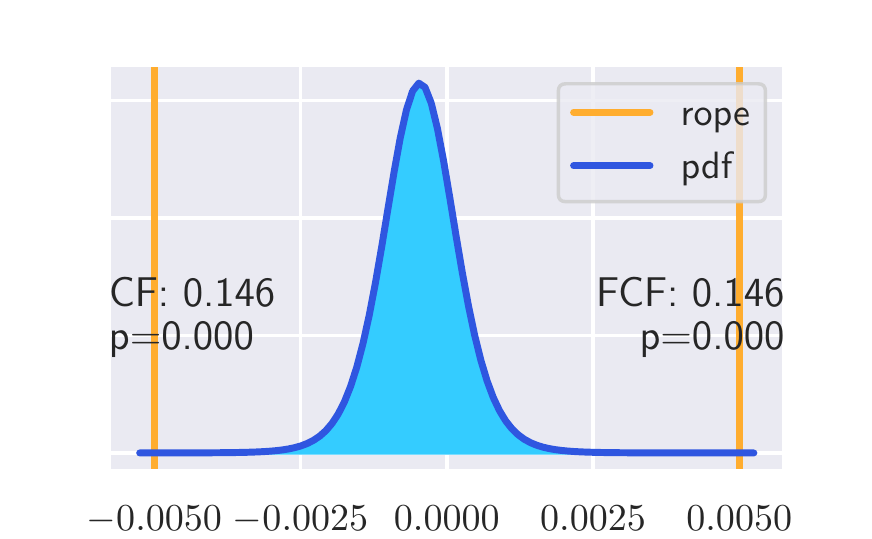}
	\end{minipage}
	\begin{minipage}{.19\textwidth}
		\centering
		\includegraphics[width=\textwidth]{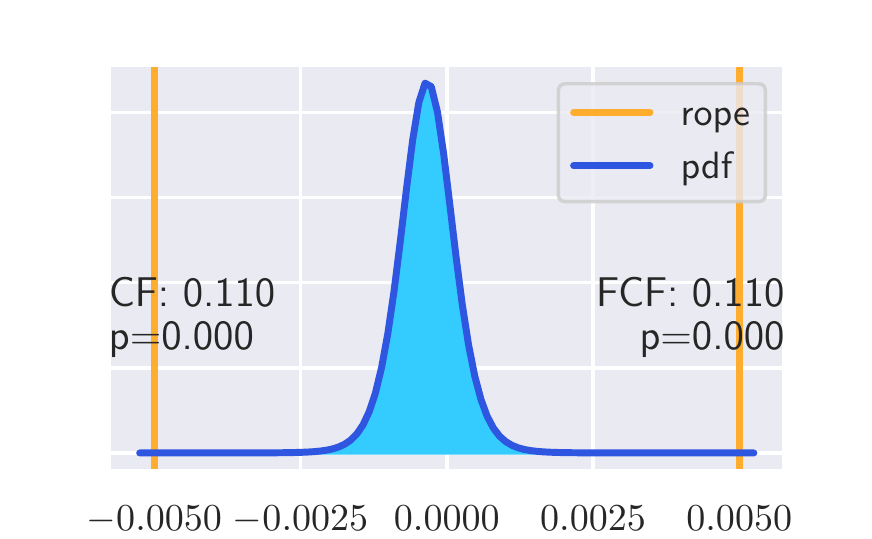}
	\end{minipage}
	\begin{minipage}{.19\textwidth}
		\centering
		\includegraphics[width=\textwidth]{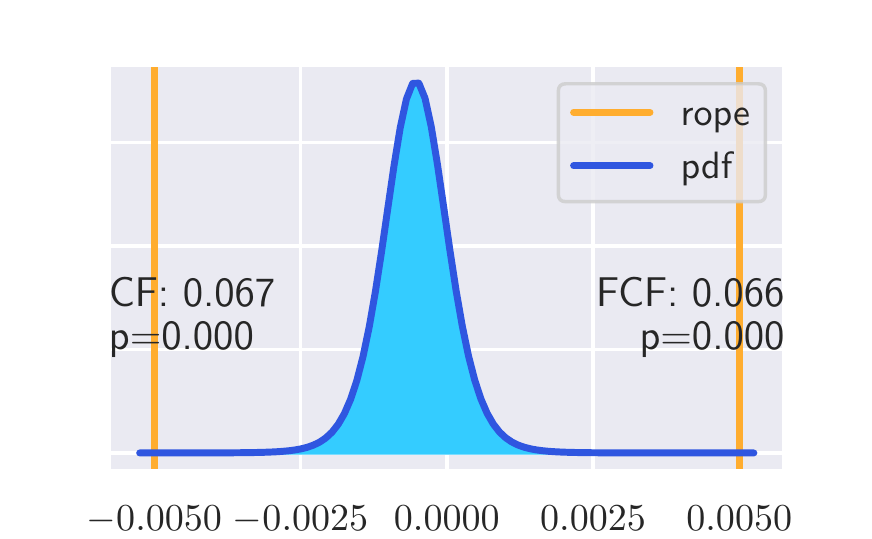}
	\end{minipage}
	\begin{minipage}{.19\textwidth}
		\centering
		\includegraphics[width=\textwidth]{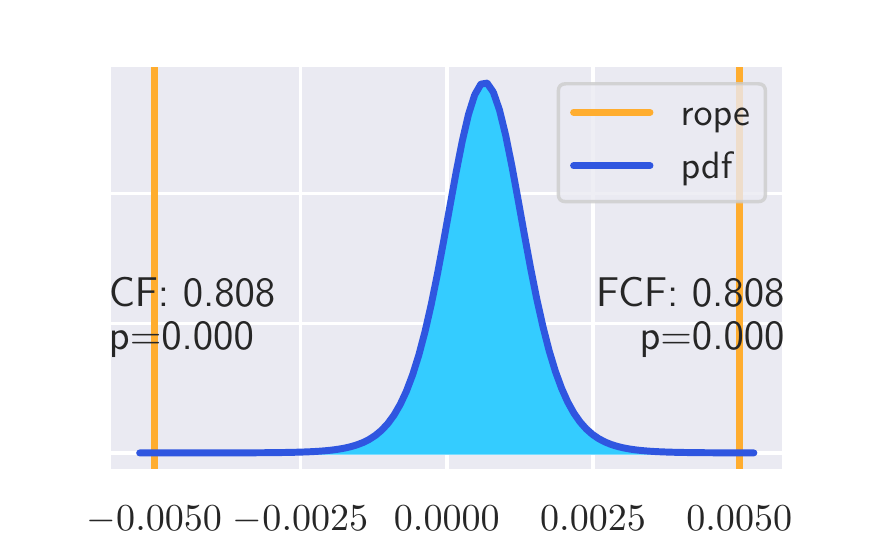}
	\end{minipage}
	\caption{Comparisons between Collaborative Filter (CF) and Federated Collaborative Filter (FCF) in form of posterior distributions drawn from a correlated Bayesian t-test, on MovieLens (top row) and in-house production (bottom row) datasets. The various performance metrics, Precision, Recall, F1, MAP and RMSE are shown in columns.  The vertical lines (rope) define a region of practical equivalence where the mean difference in performance is no more than $\pm$0.5\%. The area under this distribution in the interval [-0.005, 0.005] is 0.999 confirming that the performance of two models is statistically similar.}
	\label{fig:cf_fcf_bayes}
\end{figure*}
To further evaluate the statistical significance of the similarity between the results of CF and FCF, we performed a Bayesian correlated t-test~\cite{corani2015bayesian,benavoli2017time}. The input to the Bayesian t-test is lists of test set performance scores (for instance Precision, Recall, F1, MAP or RMSE) of the two models, obtained from 10 rebuilds. The test outputs a posterior distribution that describes the mean difference in performance scores between the CF and FCF models. Since posterior is a probability density function, it can be integrated to infer a probability of the hypothesis (that is the mean difference in recommendation performance scores is smaller than $0.5\%$). 

Figure~\ref{fig:cf_fcf_bayes} shows the posterior distributions of the differences on MovieLens and in-house production datasets (in the case of simulated data, we show similar plots in Appendix). 
Consider, the results for evaluating the similarity in Precision values of the two models. The vertical lines (rope) define a region of practical equivalence where the difference in precision is no more than $\pm$0.5\% i.e. $[-0.005, 0.005]$. The FCF model has an average precision of 0.299 and CF 0.301. The plot shows the distribution of differences. The area under this distribution in the interval ($-\infty$, $−0.005$) is $9.888e-07$, which represents the probability of CF being better than FCF. Similarly, the area in the interval ($0.005$, $\infty$) equals $1.000e-04$ and denotes the probability that FCF being better than CF. The area between $[-0.005, 0.005]$, the defined region of practical equivalence (rope), is $0.999$, which is interpreted as probability of being equivalent under then $\pm$0.5\% difference threshold. The results confirm that the precision values of CF and FCF are equivalent with a probability of 0.999. Moreover, all probability density functions in Figure~\ref{fig:cf_fcf_bayes} have more than 99\% of the area within the rope, therefore, with no loss of generality, it can be claimed that the FCF and CF recommendation performance is statistically similar.

\section{CONCLUSION}

We introduced the first federated collaborative filtering method for privacy-preserving personalized recommendations. We presented the algorithm to federate the standard collaborative filter using stochastic gradient descent based approach. The convergence analysis demonstrated that the federated model achieves robust and stable solutions by incorporating an adaptive learning rate.
The empirical evidence proved that the federated collaborative filter achieves statistically similar recommendation performance compared to the standard method. 
As a practical outcome, the results establish that the federated model can provide similar quality of recommendations as the widely used standard collaborative filter while fully preserving the user's privacy.

\paragraph{\textbf{Future Work:}} We consider this work as a first step towards privacy-preserving recommendation systems with a federated model. However, we aim to explore multiple directions in this line of research in the future. A simulator-based analysis of the real-world scenario with updates arriving from clients in a continuously asynchronous fashion (online learning) could help benchmark the system. In addition, analysis on the communication payloads and efficiency could help evaluate the other practical aspects of such systems. Another important focus area for future studies is security in federated models. We intend to investigate the effects of attacks and threats by incorporating recently proposed methods~\cite{DBLP:journals/corr/abs-1807-00459,DBLP:journals/corr/abs-1808-04866,DBLP:journals/corr/abs-1811-12470} for secure federated learning.
\cleardoublepage
\newpage
\bibliographystyle{unsrt}
\bibliography{ref}

%
%
%

\end{document}